\PassOptionsToPackage{expansion=true}{microtype}
\documentclass[sigconf]{acmart}
\setcopyright{acmcopyright}
\copyrightyear{2024}
\acmConference[USC CSCI 599]{Deep Learning Final Project}{December 2025}{Los Angeles, CA, USA}
\usepackage{booktabs}
\usepackage{multirow}
\usepackage{amsfonts}
\usepackage{amsmath}
\usepackage{nicefrac}
\usepackage{xcolor}
\usepackage{graphicx}
\usepackage{algorithm}
\usepackage{algorithmic}
\usepackage{float}
\usepackage{tikz}
\usetikzlibrary{shapes,arrows,positioning,shadows,calc}

\title{VLM4Rec: Multimodal Semantic Representation for Recommendation with Large Vision-Language Models}

\author{Ty Valencia}
\affiliation{%
  \institution{University of Southern California}
  \city{Los Angeles}
  \country{USA}
}
\email{tsvalenc@usc.edu}

\author{Burak Barlas}
\affiliation{%
  \institution{University of Southern California}
  \city{Los Angeles}
  \country{USA}
}
\email{burakbar@usc.edu}

\author{Varun Singhal}
\affiliation{%
  \institution{University of Southern California}
  \city{Los Angeles}
  \country{USA}
}
\email{vrsingha@usc.edu}

\author{Ruchir Bhatia}
\affiliation{%
  \institution{University of Southern California}
  \city{Los Angeles}
  \country{USA}
}
\email{ruchirbh@usc.edu}

\author{Wei Yang}
\affiliation{%
  \institution{University of Southern California}
  \city{Los Angeles}
  \country{USA}
}
\email{wyang930@usc.edu}

\begin{document}
\sloppy
\raggedbottom
\begin{abstract}
Multimodal recommendation is commonly framed as a feature fusion problem, where textual and visual signals are combined to better model user preference.
However, the effectiveness of multimodal recommendation may depend not only on how modalities are fused, but also on whether item content is represented in a semantic space aligned with preference matching.
This issue is particularly important because raw visual features often preserve appearance similarity, while user decisions are typically driven by higher-level semantic factors such as style, material, and usage context.
Motivated by this observation, we propose \textbf{LVLM-grounded Multimodal Semantic Representation for Recommendation (VLM4Rec)}, a lightweight framework that organizes multimodal item content through semantic alignment rather than direct feature fusion.
\textbf{VLM4Rec} first uses a large vision--language model to ground each item image into an explicit natural-language description, and then encodes the grounded semantics into dense item representations for preference-oriented retrieval.
Recommendation is subsequently performed through a simple profile-based semantic matching mechanism over historical item embeddings, yielding a practical offline--online decomposition.
Extensive experiments on multiple multimodal recommendation datasets show that \textbf{VLM4Rec} consistently improves performance over raw visual features and several fusion-based alternatives, suggesting that representation quality may matter more than fusion complexity in this setting.
The code is released at \url{https://github.com/tyvalencia/enhancing-mm-rec-sys}.
\end{abstract}



\maketitle
\title[Beyond Fusion with LVLM-grounded Semantic Representation]{VLM4Rec: Multimodal Semantic Representation for Recommendation with Large Vision-Language Models}

\section{Introduction}
\label{sec:introduction}

Recommender systems play a central role in modern e-commerce and content platforms, where users must navigate increasingly large item catalogs under limited attention~\cite{kan2026conflating,zhao2025hierarchical,yang2025structured}.
In many important application domains, including fashion, consumer goods, and lifestyle products, recommendation is inherently multimodal: items are presented not only through sparse textual metadata, but also through rich visual content that conveys appearance, material, style, and usage context.
These signals are often essential to user preference formation~\cite{wang2024rethinking, li2023text,yang2023multimodal}.
For example, two products with nearly identical titles may differ substantially in aesthetic style or occasion suitability, while two visually distinct items may still satisfy similar user needs at the semantic level.
This has made multimodal recommendation an important and rapidly growing research direction, with the goal of leveraging both textual and visual item information to better capture user preference.

Most existing multimodal recommendation methods are built around a fusion-centered paradigm~\cite{wang2021dualgnn,yang2023modal,jiang2024diffmm,li2026recgoat}.
Starting from modality-specific representations, they seek to combine textual and visual signals through concatenation, averaging, attention, gating, graph propagation, or more elaborate spectral and multi-view architectures~\cite{yang2025fitmm,zhang2024smore, ong2024spectrum}.
Within this paradigm, the main modeling question is typically how to fuse heterogeneous modalities effectively so that complementary information can be exploited downstream.
This view is natural and has motivated a broad family of effective methods.
Indeed, our own initial exploration also followed this line by comparing text-only, vision-only, and multiple fusion-based strategies within a unified retrieval framework.
Yet this perspective leaves open a more fundamental question: are the native modality spaces themselves the most suitable interface for recommendation?

A closer look suggests that the answer may not always be yes.
Raw visual features are typically optimized to preserve appearance-level similarity, whereas recommendation often depends on higher-level semantic concepts such as style, material, functionality, seasonality, or occasion.
As a result, two items may be close in visual feature space while being far apart in preference semantics; conversely, semantically substitutable items may differ substantially in low-level visual characteristics.
Short titles present a complementary challenge: although they are easy to encode, they are often too sparse to expose the semantic factors that actually drive user decisions.
Taken together, these observations suggest that the central challenge in multimodal recommendation may not be only how to combine modalities, but also how to express multimodal item content in a representation space that is compatible with preference matching~\cite{ren2024enhancing, liao2024llara, wei2024llmrec}.

This perspective also reframes how one should interpret the role of architectural complexity.
If a stronger multimodal recommender performs better, it is not always clear whether the improvement comes from a more effective fusion mechanism or simply from a better underlying representation.
Our empirical findings point strongly toward the latter explanation.
Across multiple fusion strategies, representation quality emerges as the dominant factor, substantially outweighing architectural choice.
In particular, text-only item representations derived from LLaVA-generated visual descriptions outperform all evaluated fusion variants on the LLaVA-covered subset, including attention-based fusion and SMORE-style spectral fusion.
This pattern suggests that, at least in our setting, the quality of the semantic representation may matter more than the sophistication of the fusion architecture itself.
It therefore becomes natural to view multimodal recommendation not only as a fusion problem, but also as a \emph{semantic alignment} problem~\cite{wang2024recgpt4v, liu2024rec,wang2024mllm4rec}.

At the same time, adopting such a perspective raises an important practical question.
Large vision--language models (LVLMs) provide a powerful mechanism for extracting rich semantics from images, but direct online deployment is expensive and difficult to scale in real-world recommendation systems.
Thus, even if LVLMs can provide a more meaningful semantic interface for item understanding, an effective recommendation framework must incorporate them in a way that preserves these semantic benefits without incurring prohibitive latency at serving time~\cite{zhou2025large,liu2024multimodal, shen2024pmg}.
This consideration motivates an offline--online decomposition in which expensive semantic grounding is performed once offline, while online recommendation remains a lightweight retrieval problem.
Such a design is also well aligned with our implementation, where LLaVA-NeXT 7B is used to generate item descriptions offline and Sentence-BERT is then used to encode them for efficient retrieval.

Motivated by these observations, we propose a lightweight multimodal recommendation framework built around semantic alignment rather than direct feature fusion.
At the core of our approach is \emph{LVLM-grounded Multimodal Semantic Representations for Recommendation} (\textbf{VLM4Rec}), a paradigm that uses a large vision--language model to ground each item image into an explicit natural-language description and then organizes the resulting semantics in a space more compatible with user preference matching.
Concretely, our framework first applies an LVLM to transform visual evidence into semantically interpretable content, and then encodes the grounded descriptions into dense item representations for retrieval.
Recommendation is subsequently performed through a simple profile-based semantic matching mechanism that averages historical item embeddings and ranks candidate items by cosine similarity.
The downstream design is intentionally lightweight. By avoiding a highly expressive recommendation head, we can more cleanly isolate the contribution of the item representation itself.
In this way, the proposed framework serves not only as a practical recommendation pipeline, but also as a concrete instantiation of a broader hypothesis: LVLM-grounded VLM4Rec can provide a more effective interface for multimodal recommendation than increasingly sophisticated fusion architectures.
In summary, the main contributions of this paper can be summarized as follows:
\begin{itemize}
    \item We introduce a semantic-alignment perspective on multimodal recommendation, arguing that the problem should be understood not only in terms of modality fusion, but also in terms of whether multimodal content is represented in a space compatible with preference matching.
    \item We propose a lightweight framework \textbf{VLM4Rec} that instantiates this perspective through LVLM-based visual semantic grounding, preference-aligned semantic representation, and efficient semantic retrieval.
    \item Extensive experiments on multimodal recommendation settings demonstrate that LVLM-grounded item representations can outperform raw visual embeddings and several fusion-based alternatives, suggesting that representation quality may be a more important driver of recommendation performance than fusion complexity in our setting.
\end{itemize}

\section{Related Work}

\subsection{Multimodal Recommendation}

Multimodal recommendation has evolved from incorporating content as auxiliary side information to developing architectures that explicitly model cross-modal interactions~\cite{yang2023multimodal, wang2024rethinking, li2023text}.
Early methods primarily addressed data sparsity by injecting pretrained visual or textual features into collaborative filtering pipelines.
Representative examples include VBPR~\cite{he2016vbpr}, visually aware recommendation models~\cite{kang2017visually}, and personalized multimedia recommendation frameworks~\cite{chen2019personalized}, which demonstrated that visual signals can complement ID-based recommendation, especially when interaction data are limited.
Subsequent work moved beyond simple feature augmentation and began modeling richer user--item--content dependencies through graph-based learning.
Examples include MMGCN~\cite{wang2021mmgcn, wei2019mmgcn}, graph-based multimodal recommenders~\cite{wei2020graph}, latent semantic graph construction~\cite{zhang2021mining}, and intention-aware graph architectures such as DualGNN~\cite{wang2021dualgnn}.
More recent advances further extend this line with robust diffusion-based modeling~\cite{jiang2024diffmm}, reflecting a broader trend toward more expressive multimodal interaction modeling.

Alongside this architectural evolution, an increasingly important research question concerns how to align and fuse heterogeneous modalities effectively~\cite{yang2023modal,tao2022self,liang2022semantic}.
A growing body of work has shown that naive fusion can introduce redundancy, amplify modality-specific noise, or even degrade recommendation quality when modalities are weakly aligned.
To address this issue, recent methods have explored hierarchical or alignment-aware strategies, including AlignRec~\cite{liu2024alignrec}, spectral filtering approaches such as SMORE~\cite{zhang2024smore, ong2024spectrum,yang2025structured}, and adaptive weighting or teacher-guided transfer mechanisms such as PromptMM~\cite{wei2024promptmm}.
Contrastive learning has also emerged as an important tool for multimodal representation learning, with methods based on augmented views or modality-invariant objectives improving robustness to noise and cross-modal inconsistency~\cite{yi2022multi, zhou2023bootstrap}.
Related efforts have been studied across short-video recommendation~\cite{han2022modality}, music recommendation~\cite{xu2023musenet}, and social recommendation settings~\cite{yu2022graph}, as well as in alignment-sensitive and semantics-aware frameworks~\cite{yang2023based, yang2024multimodal, guo2022topicvae}.
Despite these advances, most prior work still treats multimodal recommendation primarily as a fusion problem in native modality feature spaces.
In contrast, our work is motivated by a different question: whether multimodal item content should first be organized into a semantically explicit, preference-aligned representation space before downstream matching.

\subsection{Large Language Models and Vision--Language Models for Recommendation}
Recent progress in large language models (LLMs)~\cite{naveed2025comprehensive,chen2025tourrank,ping2025hdlcore,yang2024enhancing,yang2025maestro} has reshaped the foundation of modern AI systems. 
Beyond improving core capabilities, recent studies have also investigated how LLMs can be better aligned, adapted, and deployed across diverse settings~\cite{chang2025survey,yang2025learning,chen2026self,yang2025toward}. 
These developments have introduced new paradigms for a variety of downstream tasks~\cite{zhao2023survey,li2025climatellm,ye2024domain,yang2026auditing}. Early studies such as P5~\cite{geng2022recommendation} showed that diverse recommendation tasks can be unified under a pretrain--prompt--predict formulation, while TALLRec~\cite{bao2023tallrec} demonstrated that instruction tuning can adapt general-purpose LLMs to recommendation objectives.
Subsequent work further explored LLM-based recommendation from multiple angles, including enhanced reasoning over user history~\cite{ren2024enhancing, liao2024llara, wei2024llmrec}, recommendation-oriented pretraining~\cite{yue2023llamarec, wang2023recmind, zhang2023recommendation}, semantic tokenization~\cite{zhu2024cost}, adapter-based tuning~\cite{geng2023vip5}, and modular fine-tuning schemes~\cite{liu2024once}.
These efforts suggest that LLMs can provide a more unified expressive interface for recommendation than conventional shallow feature fusion~\cite{gu2025r,xia2025hierarchical,xia2025trackrec}.

Large vision--language models (LVLMs) extend this paradigm by adding direct visual understanding.
Models such as GPT-4V and LLaVA enable joint reasoning over images and text, making them especially attractive for multimodal recommendation scenarios where visual semantics are central to user preference.
Several recent studies have begun exploring this direction.
For example, Rec-GPT4V~\cite{wang2024recgpt4v, liu2024rec} uses visual-text prompting to infer user preferences from product images, while MLLM4Rec~\cite{wang2024mllm4rec} and MM-REACT~\cite{yang2023mm} investigate how multimodal LLM-style systems can support recommendation-oriented reasoning.
Other works have explored unified multimodal recommendation with foundation models, including VIP5~\cite{geng2023vip5}, MMRec~\cite{tian2024mmrec}, and related studies on when and how multimodal large models truly benefit recommendation accuracy~\cite{zhou2025large}.
Beyond retrieval and ranking, LLM-based multimodal methods have also been extended to generation-oriented applications such as personalized news, ad copy, and visual narratives~\cite{liu2024multimodal, shen2024pmg}.

Our work is closely related to this emerging line, but differs in both emphasis and design.
Rather than using LVLMs as end-to-end online recommenders or prompting engines, we employ them as semantic grounding modules that transform item images into explicit natural-language descriptions.
These grounded descriptions are then encoded into \emph{LVLM-grounded Multimodal Semantic Representations for Recommendation} (VLM4Rec), which serve as the basis for lightweight retrieval.
In this sense, our method is less focused on increasing reasoning complexity at inference time and more focused on constructing a semantically aligned item representation space that is both effective and practical for recommendation.

\section{Method}
\label{sec:method}

\subsection{Problem Formulation}
\label{sec:problem}

We study top-$K$ recommendation under implicit feedback.
Let $\mathcal{U}$ and $\mathcal{I}$ denote the user set and item set, respectively.
The observed interaction data are given by
\begin{equation}
\mathcal{D} = \{(u,i)\mid u\in\mathcal{U},\, i\in\mathcal{I}\},
\end{equation}
where $(u,i)\in\mathcal{D}$ indicates that user $u$ has interacted with item $i$.
We partition the interaction set into disjoint training and evaluation subsets, denoted by $\mathcal{D}_{\mathrm{train}}$ and $\mathcal{D}_{\mathrm{eval}}$.

Each item $i\in\mathcal{I}$ is associated with multimodal content.
In our setting, this includes an image $x_i$ and, when available, a title or short textual metadata $t_i$.
For each user $u$, let
\begin{equation}
\mathcal{N}_u = \{i\in\mathcal{I}\mid (u,i)\in\mathcal{D}_{\mathrm{train}}\}
\end{equation}
denote the set of items appearing in the user's training interactions.
Following the practical setting adopted in our system, we retain only the most recent $L_{\max}$ interactions and define the user history as
\begin{equation}
\mathcal{H}_u = \mathrm{Last}(\mathcal{N}_u, L_{\max}),
\qquad L_{\max}=10.
\end{equation}

Our objective is to construct an item representation space in which semantically preference-relevant items are close to one another, so that a user's historical interactions can be used to retrieve future items of interest.
Given $\mathcal{H}_u$, we rank all candidate items $j\in\mathcal{I}\setminus\mathcal{H}_u$ and return a top-$K$ recommendation set
\begin{equation}
\hat{\mathcal{R}}_u^K \subseteq \mathcal{I}\setminus\mathcal{H}_u.
\end{equation}

Unlike much of the multimodal recommendation literature, our emphasis is not on designing an increasingly expressive fusion architecture.
Instead, we investigate whether recommendation quality depends more fundamentally on how multimodal item content is organized into a semantic space aligned with downstream preference matching.
This perspective leads to a lightweight yet conceptually focused framework consisting of three stages: visual semantic grounding, preference-aligned semantic representation, and semantic matching.

\subsection{Framework Overview}
\label{sec:overview}

The central premise of our framework is that multimodal recommendation should not be viewed solely as a feature fusion problem.
For recommendation, the key issue is not merely preserving modality-specific information, but representing heterogeneous item content in a space that is compatible with how users compare, interpret, and choose items.
In particular, raw visual embeddings may capture rich appearance information, yet remain only weakly aligned with semantic concepts that often govern user decisions, such as style, material, usage scenario, seasonality, or perceived functionality.

To instantiate this perspective, we propose a three-stage framework:
\begin{enumerate}
    \item \textbf{Visual Semantic Grounding}: transform each item image into an explicit semantic description using a large vision--language model;
    \item \textbf{Preference-Aligned Semantic Representation}: map grounded descriptions into a dense semantic embedding space for retrieval;
    \item \textbf{Semantic Matching}: construct user profiles by aggregating historical item embeddings and retrieve candidate items via similarity search.
\end{enumerate}

\begin{figure*}[t]
  \centering
  \includegraphics[width=\textwidth]{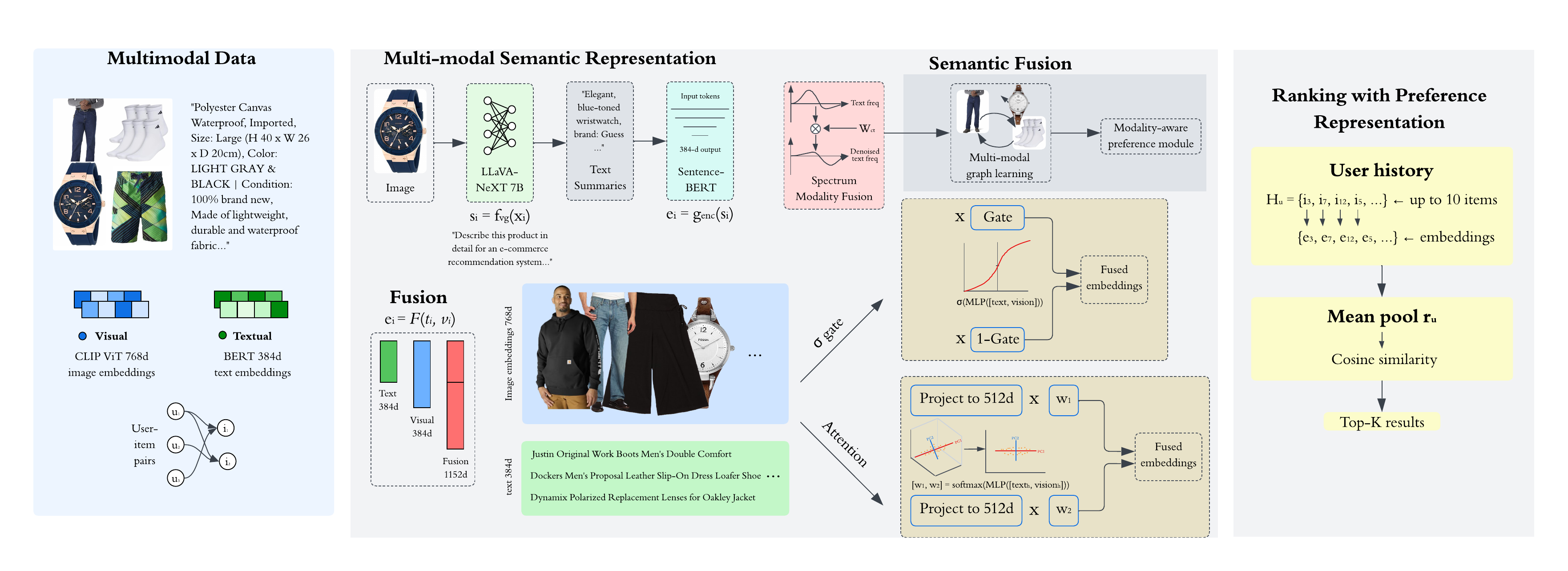}
  \caption{Overview of the VLM4Rec framework. Multimodal item content (images, titles, and user--item interactions) is processed through multiple representation paths: the proposed LVLM grounding path (top), which transforms item images into rich semantic descriptions via LLaVA-NeXT 7B and encodes them with Sentence-BERT. Baseline fusion and learned combination methods including concatenation, gating, and attention (middle), and SMORE spectral fusion (top). All embedding variants feed into the same lightweight retrieval stage, where user profiles are constructed via mean pooling and candidates are ranked by cosine similarity.}
  \label{fig:model-overview}
\end{figure*}

Formally, for each item $i$, the framework constructs an embedding
\begin{equation}
\mathbf{e}_i = g_{\mathrm{enc}}\!\left(f_{\mathrm{vg}}(x_i)\right) \in \mathbb{R}^{d},
\label{eq:full_pipeline}
\end{equation}
where $f_{\mathrm{vg}}(\cdot)$ denotes the visual grounding module and $g_{\mathrm{enc}}(\cdot)$ denotes the semantic encoder.
For a user $u$, the user profile is computed from the embeddings of items in $\mathcal{H}_u$, and recommendation is then performed in the resulting semantic space.

The downstream design is intentionally simple.
By avoiding a highly expressive matching model, we are able to isolate the contribution of the item representations themselves.
Consequently, if gains are observed under such a lightweight retriever, they can be more credibly attributed to the semantic quality of the item embeddings rather than to the capacity of a complex recommendation head.

\subsection{Visual Semantic Grounding}
\label{sec:grounding}

In multimodal recommendation, image features are typically consumed either directly as visual embeddings or indirectly through a fusion module.
However, raw visual representations are primarily optimized to preserve visual similarity rather than to expose the semantic factors that matter for user preference.
For instance, two items may be visually similar in texture or silhouette while differing substantially in material, usage context, or suitability for a particular occasion.
Conversely, items that differ in low-level appearance may still be semantically similar from a recommendation standpoint.
This mismatch motivates us to first ground visual evidence into explicit semantics.

\paragraph{Grounding Function.}
For each item image $x_i$, we generate a natural-language description
\begin{equation}
s_i = f_{\mathrm{vg}}(x_i),
\label{eq:grounding}
\end{equation}
where $f_{\mathrm{vg}}$ is instantiated with a large vision--language model.
In our implementation, we use LLaVA-NeXT 7B to produce fine-grained textual descriptions of product images.
The prompting strategy encourages the model to describe recommendation-relevant attributes such as color, material, style, category cues, and likely usage scenarios.

The output $s_i$ serves as a semantically explicit interface for the visual modality.
Rather than retaining image information solely in a latent visual feature space, the LVLM transforms it into a structured linguistic representation that is naturally amenable to semantic comparison.
This is particularly desirable in recommendation, where user preferences are often expressed, implicitly or explicitly, in semantic rather than low-level visual terms.

\paragraph{Semantic Interpretation.}
Conceptually, Eq.~\eqref{eq:grounding} is not merely an image captioning step.
Instead, it performs a task-oriented semantic abstraction:
\begin{equation}
x_i \longrightarrow s_i \approx \text{recommendation-relevant semantic factors}.
\end{equation}
The resulting description may encode multiple factors simultaneously, such as appearance, material, style, occasion, and usage context.
Such compositional semantics are difficult to enforce directly in raw visual embeddings, whereas they are naturally expressible in language.

\paragraph{Offline Semantic Cache.}
An important practical property of this stage is that it can be executed entirely offline.
Let $\mathcal{I}_{\mathrm{cov}}\subseteq\mathcal{I}$ denote the subset of items for which grounded descriptions are generated.
The grounding stage then produces a semantic cache
\begin{equation}
\mathcal{S} = \{(i,s_i)\mid i\in\mathcal{I}_{\mathrm{cov}}\},
\end{equation}
which can be reused across all users during retrieval.
This decouples the expensive LVLM inference from online recommendation and turns semantic grounding into a one-time preprocessing step rather than a runtime cost.

\subsection{Preference-Aligned Semantic Representation}
\label{sec:encoding}

Once grounded descriptions are obtained, we map them into a dense semantic embedding space using a text encoder:
\begin{equation}
\mathbf{e}_i = g_{\mathrm{enc}}(s_i), \qquad \mathbf{e}_i \in \mathbb{R}^{d}.
\label{eq:item_embedding}
\end{equation}
In our implementation, $g_{\mathrm{enc}}$ is instantiated with Sentence-BERT (all-MiniLM-L6-v2), yielding $d=384$-dimensional sentence embeddings.

Unlike conventional text features derived only from short titles, the description $s_i$ already provides a semantically enriched view of the visual content.
Accordingly, the encoder in Eq.~\eqref{eq:item_embedding} does more than compress text into a vector.
Rather, it induces a \emph{preference-aligned semantic space} in which image-derived item semantics can be compared through standard vector similarity.

\paragraph{Preference Alignment Perspective.}
Let $\phi(i)$ denote the latent set of recommendation-relevant semantic factors associated with item $i$.
Although $\phi(i)$ is not directly observable, the grounding and encoding pipeline aims to produce
\begin{equation}
\mathbf{e}_i \approx h(\phi(i)),
\end{equation}
for some smooth embedding map $h(\cdot)$.
Under this view, the role of the representation is to make similarity in embedding space better reflect similarity in preference semantics:
\begin{equation}
\mathrm{sim}(\mathbf{e}_i,\mathbf{e}_j)
\propto
\mathrm{overlap}\!\left(\phi(i),\phi(j)\right).
\label{eq:semantic_overlap}
\end{equation}
This is precisely the property desired for retrieval-based recommendation.
Items need not be identical in raw appearance; rather, they should be close in the semantic dimensions that matter for user choice.

\paragraph{Unified Semantic Space.}
A key distinction between our framework and direct multimodal fusion lies in where alignment occurs.
In conventional fusion methods,
\begin{equation}
\mathbf{e}_i^{\mathrm{fusion}} = F\!\left(\mathbf{v}_i, \mathbf{t}_i\right),
\end{equation}
where $\mathbf{v}_i$ and $\mathbf{t}_i$ are native visual and textual features and $F$ is a fusion operator such as concatenation, averaging, attention, or gating.
In contrast, our framework first transforms image content into a semantically explicit modality and then encodes it into a unified semantic space:
\begin{equation}
\mathbf{e}_i = g_{\mathrm{enc}}\!\left(f_{\mathrm{vg}}(x_i)\right).
\end{equation}
This distinction is conceptually important.
Rather than asking a fusion module to reconcile heterogeneous feature spaces, we ask the LVLM and the text encoder to jointly construct a space in which semantic comparability is built in from the outset.

\subsection{Semantic Matching for Recommendation}
\label{sec:matching}

Given normalized item embeddings, we represent each user by aggregating the embeddings of items in the user's recent interaction history.
For item normalization, we first define
\begin{equation}
\tilde{\mathbf{e}}_i = \frac{\mathbf{e}_i}{\|\mathbf{e}_i\|_2}.
\end{equation}

\paragraph{User Profile Construction.}
For user $u$, the unnormalized profile is defined as
\begin{equation}
\mathbf{r}_u =
\frac{1}{|\mathcal{H}_u|}
\sum_{i\in\mathcal{H}_u}
\tilde{\mathbf{e}}_i.
\label{eq:user_profile_raw}
\end{equation}
We further normalize the user profile:
\begin{equation}
\tilde{\mathbf{r}}_u =
\frac{\mathbf{r}_u}{\|\mathbf{r}_u\|_2}.
\label{eq:user_profile_norm}
\end{equation}

This mean-pooling estimator may be interpreted as the empirical preference center of the user in the aligned semantic space.
Although simple, it is well matched to our objective: we seek to examine whether semantically grounded item embeddings already make user preference geometry easier to recover, even without a sophisticated sequential encoder.

\paragraph{Candidate Scoring.}
For each candidate item $j\in\mathcal{I}\setminus\mathcal{H}_u$, we compute the matching score using cosine similarity:
\begin{equation}
\mathrm{score}(u,j)
=
\cos(\tilde{\mathbf{r}}_u,\tilde{\mathbf{e}}_j)
=
\tilde{\mathbf{r}}_u^\top \tilde{\mathbf{e}}_j.
\label{eq:cosine_score}
\end{equation}
Since both vectors are normalized, the score reduces to an inner product.
The final top-$K$ recommendation list is
\begin{equation}
\hat{\mathcal{R}}_u^K
=
\mathrm{TopK}_K
\left(
\left\{
(j,\mathrm{score}(u,j))
\mid
j\in\mathcal{I}\setminus\mathcal{H}_u
\right\}
\right).
\label{eq:topk}
\end{equation}

\paragraph{Matrix Form.}
For efficient retrieval, all normalized item embeddings can be stacked into
\begin{equation}
\tilde{\mathbf{E}} =
\begin{bmatrix}
\tilde{\mathbf{e}}_1^\top \\
\tilde{\mathbf{e}}_2^\top \\
\vdots \\
\tilde{\mathbf{e}}_{|\mathcal{I}|}^\top
\end{bmatrix}
\in \mathbb{R}^{|\mathcal{I}|\times d}.
\end{equation}
Then, for a fixed user $u$, the score vector over all items is
\begin{equation}
\mathbf{s}_u = \tilde{\mathbf{E}}\,\tilde{\mathbf{r}}_u \in \mathbb{R}^{|\mathcal{I}|}.
\label{eq:score_vector}
\end{equation}
After masking items in $\mathcal{H}_u$, recommendation reduces to selecting the indices of the largest $K$ entries in $\mathbf{s}_u$.
This formulation makes the retrieval stage compatible with efficient nearest-neighbor search or vector database deployment.

\paragraph{Rationale for a Lightweight Retriever.}
A natural question is why we do not employ a more expressive user encoder or ranking network.
The answer is methodological.
If a powerful downstream model were introduced, empirical improvements could arise either from better item semantics or from increased ranking capacity.
By deliberately adopting the simple matching rule in Eqs.~\eqref{eq:user_profile_raw}--\eqref{eq:topk}, we isolate the representational contribution of the semantic alignment pipeline itself.

\subsection{Optimization and Training}
\label{sec:opt}

A notable property of our framework is that the core semantic transformation is achieved by pretrained modules rather than by task-specific end-to-end optimization.
Specifically, both the visual semantic grounding module $f_{\mathrm{vg}}(\cdot)$ and the semantic encoder $g_{\mathrm{enc}}(\cdot)$ are used in a frozen manner, and the resulting item representations are precomputed offline.
Accordingly, the main pipeline does not introduce an additional trainable recommendation head; instead, it relies on the quality of the aligned semantic space itself to support downstream retrieval.

Recall that the item representation in our framework is
\begin{equation}
\mathbf{e}_i = g_{\mathrm{enc}}\!\left(f_{\mathrm{vg}}(x_i)\right),
\qquad
\tilde{\mathbf{e}}_i = \frac{\mathbf{e}_i}{\|\mathbf{e}_i\|_2},
\end{equation}
and that the user profile is constructed by mean pooling over the recent interaction history:
\begin{equation}
\mathbf{r}_u = \frac{1}{|\mathcal{H}_u|}\sum_{i\in\mathcal{H}_u}\tilde{\mathbf{e}}_i,
\qquad
\tilde{\mathbf{r}}_u = \frac{\mathbf{r}_u}{\|\mathbf{r}_u\|_2}.
\end{equation}
The final matching score for candidate item $j$ is therefore
\begin{equation}
s(u,j) = \tilde{\mathbf{r}}_u^\top \tilde{\mathbf{e}}_j.
\label{eq:opt_score}
\end{equation}

Under this formulation, our method may be understood as addressing the recommendation problem \emph{through representation construction} rather than through a separately learned scoring function.
That is, the burden of optimization is shifted upstream:
the LVLM grounds raw visual content into semantically explicit descriptions, and the sentence encoder maps those descriptions into a geometry in which simple similarity search is meaningful for preference matching.
From this perspective, the quality of recommendation depends primarily on whether the composed mapping
\begin{equation}
x_i \mapsto f_{\mathrm{vg}}(x_i) \mapsto g_{\mathrm{enc}}(f_{\mathrm{vg}}(x_i))
\end{equation}
induces a representation space in which semantically preference-compatible items lie close together.

Although the main pipeline is inference-only at recommendation time, it is still useful to characterize the induced ranking principle.
Given a user $u$, a positive target item $i^{+}$, and a negative item $i^{-}$, a desirable aligned semantic space should satisfy
\begin{equation}
s(u,i^{+}) > s(u,i^{-}).
\label{eq:desired_rank}
\end{equation}
Equivalently, one may interpret the representation as implicitly minimizing a pairwise ranking risk of the form
\begin{equation}
\mathcal{L}_{\mathrm{align}}
=
\sum_{(u,i^{+},i^{-})}
\ell\!\left(
s(u,i^{-}) - s(u,i^{+})
\right),
\label{eq:implicit_obj}
\end{equation}
where $\ell(\cdot)$ is a monotone surrogate loss, such as a logistic or hinge-style ranking penalty.
We do not explicitly optimize Eq.~\eqref{eq:implicit_obj} in the proposed method.
Rather, our framework examines whether pretrained visual grounding and semantic encoding are sufficient to induce a representation space that approximately satisfies Eq.~\eqref{eq:desired_rank} under a lightweight retriever.

This design choice is deliberate.
If we were to append a highly expressive trainable ranking module, empirical gains could arise either from improved item semantics or simply from increased downstream model capacity.
By keeping the matching stage parameter-free, the proposed framework isolates the contribution of \emph{semantic alignment} itself.
In other words, the method asks a more focused question: once visual content has been grounded into a preference-aligned semantic space, how much recommendation signal can already be recovered by a minimal matching rule?

From an optimization standpoint, this design also yields a clean separation between offline and online computation.
The expensive stage---LVLM-based semantic grounding followed by semantic encoding---is performed once offline for the item catalog, producing a cached embedding table
\begin{equation}
\tilde{\mathbf{E}}
=
\begin{bmatrix}
\tilde{\mathbf{e}}_1^\top\\
\tilde{\mathbf{e}}_2^\top\\
\vdots\\
\tilde{\mathbf{e}}_{|\mathcal{I}|}^\top
\end{bmatrix}
\in \mathbb{R}^{|\mathcal{I}| \times d}.
\end{equation}
At serving time, recommendation reduces to user-profile construction and vector retrieval:
\begin{equation}
\mathbf{s}_u = \tilde{\mathbf{E}}\,\tilde{\mathbf{r}}_u,
\end{equation}
followed by masking previously interacted items and selecting the top-ranked candidates.
Thus, although our framework is not optimized through end-to-end gradient updates, it is nevertheless built upon a clear optimization principle: semantic structure should be injected into the item space \emph{prior} to retrieval, so that downstream recommendation can be solved with minimal additional learning.

For completeness, trainable comparison models are optimized using their standard objectives and protocols, which are described separately in the experimental setup.
This separation keeps the method section focused on the optimization philosophy of the proposed framework rather than on implementation details of external baselines.

\subsection{Discussion}
\label{sec:discussion}

Our method is built on the view that multimodal recommendation is not solely a modality fusion problem, but also a semantic alignment problem.
Within this perspective, the role of the LVLM is not merely to generate longer text, but to ground visual evidence into semantically explicit descriptions that better expose preference-relevant item factors.
Similarly, the role of the sentence encoder is not merely to compress language, but to construct a geometry in which those grounded semantics become comparable through vector similarity.
Finally, the role of the lightweight retriever is not to maximize model complexity, but to reveal whether the aligned semantic space itself is already useful for preference matching.

We intentionally avoid making an overly broad claim that this paradigm must outperform all fusion-based methods in every setting.
Our position is more modest and more precise: in our setting, the empirical evidence suggests that organizing multimodal content into a preference-aligned semantic space can be more beneficial than increasing the sophistication of feature fusion.
From this perspective, semantic alignment provides a useful lens for understanding why LVLM-grounded item representations can be effective for multimodal recommendation.

\section{Experiments}
\label{sec:experiments}

We conduct comprehensive experiments addressing five research questions: How does offline LVLM-generated text compare to traditional baselines? (RQ1) Does text-only LVLM outperform multimodal fusion? (RQ2) How does representation quality impact fusion architectures? (RQ3) Are improvements consistent across metrics and ranking depths? (RQ4) And what mechanisms explain LVLM effectiveness? (RQ5) The experiments systematically isolate the effects of representation quality and fusion architecture to understand which factors drive recommendation improvements.

\begin{table*}[!t]
  \centering
  \small
  \setlength{\tabcolsep}{5pt}
  \renewcommand{\arraystretch}{1.12}
  \begin{tabular}{lccccccc}
    \toprule
    \textbf{Model} & \textbf{Recall@5} & \textbf{Recall@10} & \textbf{Recall@20} & \textbf{NDCG@10} & \textbf{NDCG@20} & \textbf{Hit@10} & \textbf{Improv. over BERT} \\
    \midrule
    \multicolumn{8}{l}{\textit{Text-Only Models}} \\
    LLaVA-NeXT 7B      & 0.289 & \textbf{0.354} & 0.394 & 0.274 & 0.285 & 0.370 & \textbf{+54.9\%} \\
    BERT Text-Only     & 0.198 & 0.228          & 0.269 & 0.184 & 0.198 & 0.320 & baseline \\
    \midrule
    \multicolumn{8}{l}{\textit{LLaVA-based Fusion Variants}} \\
    LLaVA + Attention      & 0.256 & 0.310 & 0.367 & 0.250 & 0.265 & 0.325 & +35.8\% \\
    LLaVA + Concatenation  & 0.173 & 0.283 & 0.329 & 0.176 & 0.189 & 0.304 & +23.7\% \\
    LLaVA + Naive Avg      & 0.173 & 0.281 & 0.329 & 0.176 & 0.190 & 0.302 & +22.8\% \\
    SMORE (LLaVA)          & 0.180 & 0.273 & 0.353 & 0.174 & 0.194 & 0.292 & +19.7\% \\
    \midrule
    \multicolumn{8}{l}{\textit{BERT-based Fusion Variants}} \\
    BERT + Attention       & 0.206 & 0.235 & 0.271 & 0.197 & 0.209 & 0.321 & +2.8\% \\
    BERT + Concatenation   & 0.157 & 0.189 & 0.234 & 0.154 & 0.169 & 0.272 & -17.4\% \\
    BERT + Naive Avg       & 0.147 & 0.180 & 0.233 & 0.143 & 0.161 & 0.264 & -21.1\% \\
    SMORE (BERT)           & 0.150 & 0.188 & 0.232 & 0.132 & 0.147 & 0.277 & -17.8\% \\
    \midrule
    \multicolumn{8}{l}{\textit{Other Baselines}} \\
    Gating Fusion          & 0.095 & 0.127 & 0.136 & 0.097 & 0.100 & 0.143 & -44.5\% \\
    Vision-Only            & 0.074 & 0.105 & 0.137 & 0.073 & 0.084 & 0.157 & -53.9\% \\
    \bottomrule
  \end{tabular}
  \caption{Recommendation performance across all twelve model variants. LLaVA-NeXT 7B text-only achieves the best overall performance, reaching 0.354 Recall@10, which is a 54.9\% improvement over the BERT text-only baseline. Notably, the text-only LLaVA variant outperforms all multimodal fusion approaches, suggesting that LVLM-generated descriptions already capture rich visual semantics in a highly recommendation-friendly form.}
  \label{tab:main-results}
\end{table*}

\subsection{Experimental Settings}

\subsubsection{Datasets}

We conduct experiments on the Kaggle Multimodal Recommendation dataset for the \emph{Clothing, Shoes, and Jewelry} category.\footnote{\url{https://www.kaggle.com/competitions/multimodal-recommendation}} The dataset, derived from Amazon product reviews, comprises 23,318 users, 38,493 items, and 178,944 interactions (113,836 training, 34,380 validation, 30,728 test), with pre-computed BERT text embeddings and CLIP vision embeddings for each item. On average, each user interacts with 7.7 items in the training set. Due to computational constraints, we generated LLaVA-NeXT 7B descriptions for 4,708 items (12.2\% of the catalog) using an NVIDIA RTX 5080 GPU with 4-bit quantization, requiring approximately 8 hours for batch inference. For fair comparison, LLaVA-based models are evaluated on 2,914 users who have LLaVA-covered items in both training and validation splits.

\subsubsection{Baselines and Comparison Strategy}

We evaluate twelve embedding strategies organized into four categories to isolate the effect of representation quality and fusion mechanisms. First, we establish text-only baselines using LLaVA-NeXT 7B and BERT title embeddings. Second, we test LLaVA-based fusion variants that combine LVLM descriptions (384-dimensional) with CLIP vision embeddings (768-dimensional) through attention mechanisms, concatenation, naive averaging, and SMORE spectral fusion \cite{zhang2024smore}. Third, we replicate these same fusion strategies using BERT title embeddings instead of LVLM descriptions to isolate the effect of representation quality. Fourth, we include additional baselines: gating fusion and vision-only CLIP embeddings. This controlled design enables direct comparison of representation quality (LVLM vs. BERT) and fusion mechanism impact across all strategies.

\subsubsection{Evaluation Protocol}

Our evaluation follows a standard protocol: we construct user profiles by mean-pooling embeddings from the last 10 items in each user's training history, rank all items by cosine similarity, and measure performance against ground-truth items from the validation split. We report Recall@K, NDCG@K (using binary relevance), and Hit-Rate@K for $K \in \{5, 10, 20\}$, computing the mean across all evaluated users. BERT-based models use the same user subset as LLaVA-based models for direct comparison, ensuring all results are directly comparable on equal footing.

\subsubsection{Implementation Details}

For LLaVA-NeXT description generation, we used the lmms-lab/llava-next-7b-hf model with 4-bit quantization, requesting detailed descriptions that emphasize visual attributes, color, style, material, and appropriate product occasions. We then encoded these descriptions using Sentence-BERT (all-MiniLM-L6-v2) to produce 384-dimensional embeddings. This offline batch processing approach, requiring approximately 8 hours for 4,708 items, avoids expensive per-query inference and enables lightweight cosine similarity operations during retrieval. For learned fusion methods, we trained attention-based fusion using InfoNCE-style contrastive loss over 15 epochs, while the SMORE baseline used pairwise BPR loss with a 2-layer GCN and $k$NN graph ($k=10$) over 20 epochs. All embeddings were L2-normalized before final recommendation scoring. Across all models, we used mean pooling for user profile aggregation from the last 10 items in training history, ensuring a consistent baseline for fair comparison.

\subsection{RQ1: How Does Offline LVLM-Generated Text Compare to Traditional Baselines?}
\label{sec:rq1}

\begin{figure*}[h]
  \centering
  \includegraphics[width=0.9\textwidth]{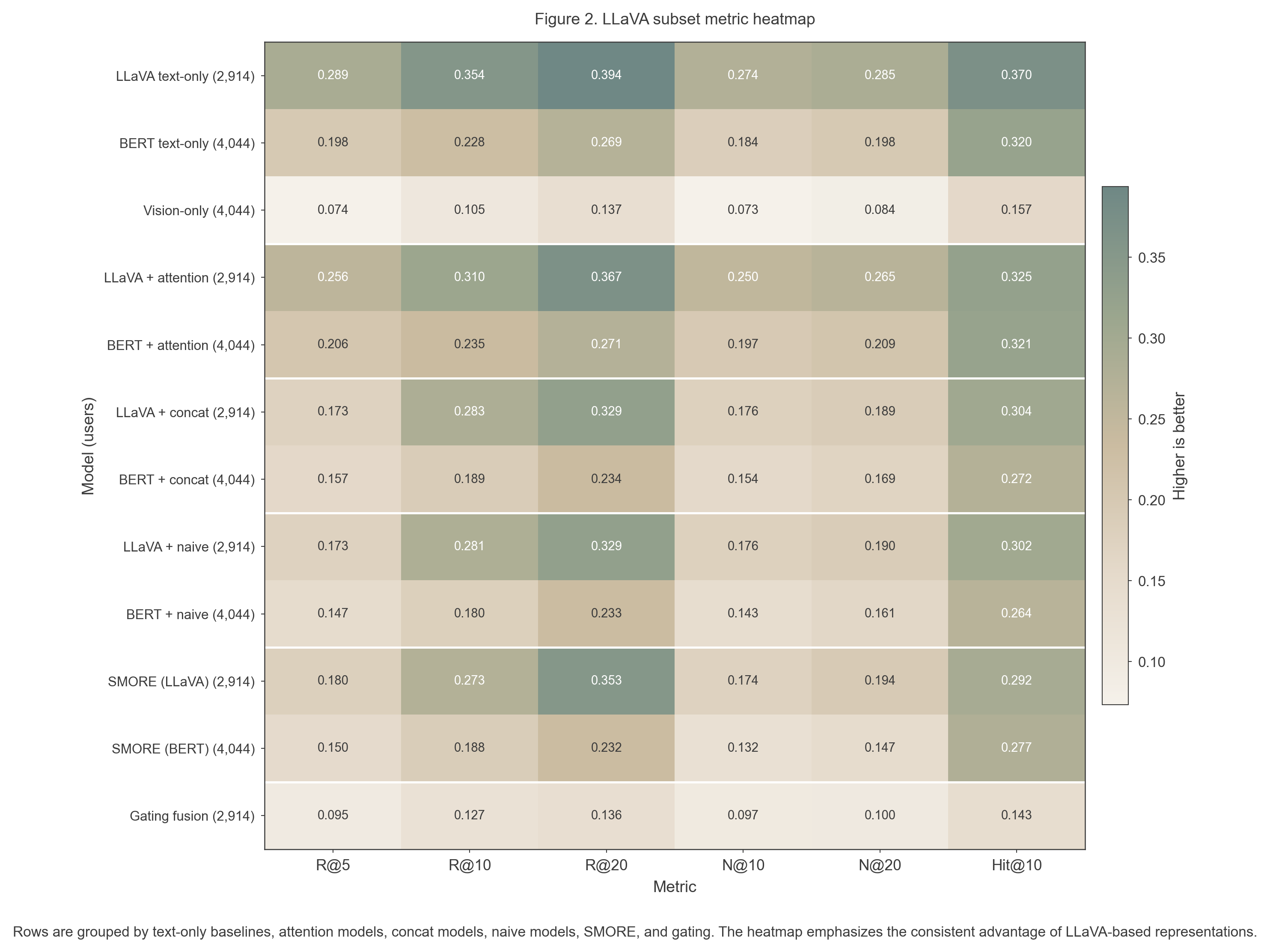}
  \caption{LLaVA-covered subset metric heatmap. Rows grouped by model type; columns show Recall@5, @10, @20, NDCG@10, @20, and Hit@10. Color intensity indicates performance (higher is better). LLaVA-based representations consistently outperform BERT-based variants across all metrics.}
  \label{fig:subset-heatmap}
\end{figure*}

We begin by comparing offline LVLM generated text representations against traditional text and vision baselines.
As shown in Table~\ref{tab:main-results}, LLaVA-NeXT 7B text-only delivers the strongest overall performance on the LLaVA-covered subset, reaching 0.354 Recall@10, compared with 0.228 for the BERT text-only baseline, corresponding to a 54.9\% relative improvement.
The gain is not confined to a single metric: LLaVA also improves Recall@5 (0.289 vs.\ 0.198), Recall@20 (0.394 vs.\ 0.269), NDCG@10 (0.274 vs.\ 0.184), and NDCG@20 (0.285 vs.\ 0.198), indicating that the advantage extends across both retrieval accuracy and ranking quality.
The heatmap in Figure~\ref{fig:subset-heatmap} further confirms this pattern, showing that LLaVA-based representations consistently dominate BERT-based counterparts across all reported metrics.
Taken together, these results provide strong evidence that offline LVLM-generated descriptions encode substantially richer recommendation-relevant semantics than conventional title-based text embeddings.

\begin{figure*}[h]
  \centering
  \includegraphics[width=0.95\textwidth]{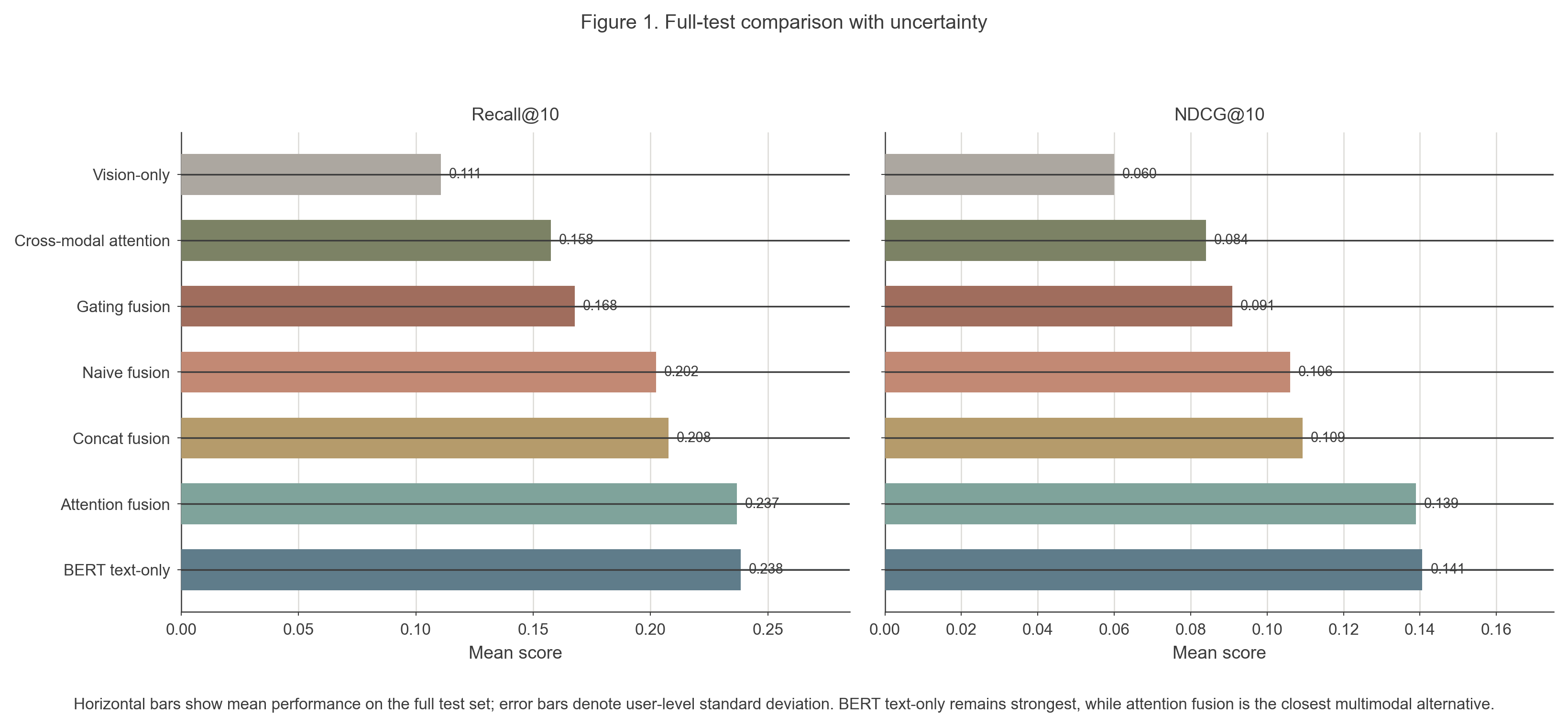}
  \caption{Full test set comparison with uncertainty. Horizontal bars show mean Recall@10 and NDCG@10 performance across all seven models evaluated on the complete dataset (23,318 users); error bars denote user-level standard deviation. BERT text-only remains strongest on the full test set, while attention fusion is the closest multimodal alternative.}
  \label{fig:full-test}
\end{figure*}

Figure~\ref{fig:full-test} offers a complementary perspective on the full test set.
Since LLaVA descriptions are only available for a subset of the catalog, the full-set comparison is not directly comparable to the matched-subset analysis above; nevertheless, it is informative as a practical reference point.
On the full test set, BERT text-only remains the strongest among the seven models that can be evaluated end-to-end over the entire item space, while attention fusion is the closest multimodal alternative.
This contrast highlights an important practical consideration: the effectiveness of LVLM-based semantic representations is clear once coverage is controlled, but realizing their full benefit at scale depends on sufficiently broad offline semantic generation.
In other words, the full-set result does not contradict the superiority of LVLM-grounded text on the matched subset; rather, it underscores the importance of coverage and scalability in deploying such representations in large recommendation catalogs.

\subsection{RQ2: Does Text-Only LVLM Outperform Multimodal Fusion?}
\label{sec:rq2}

\begin{figure*}[h]
  \centering
  \includegraphics[width=0.8\textwidth]{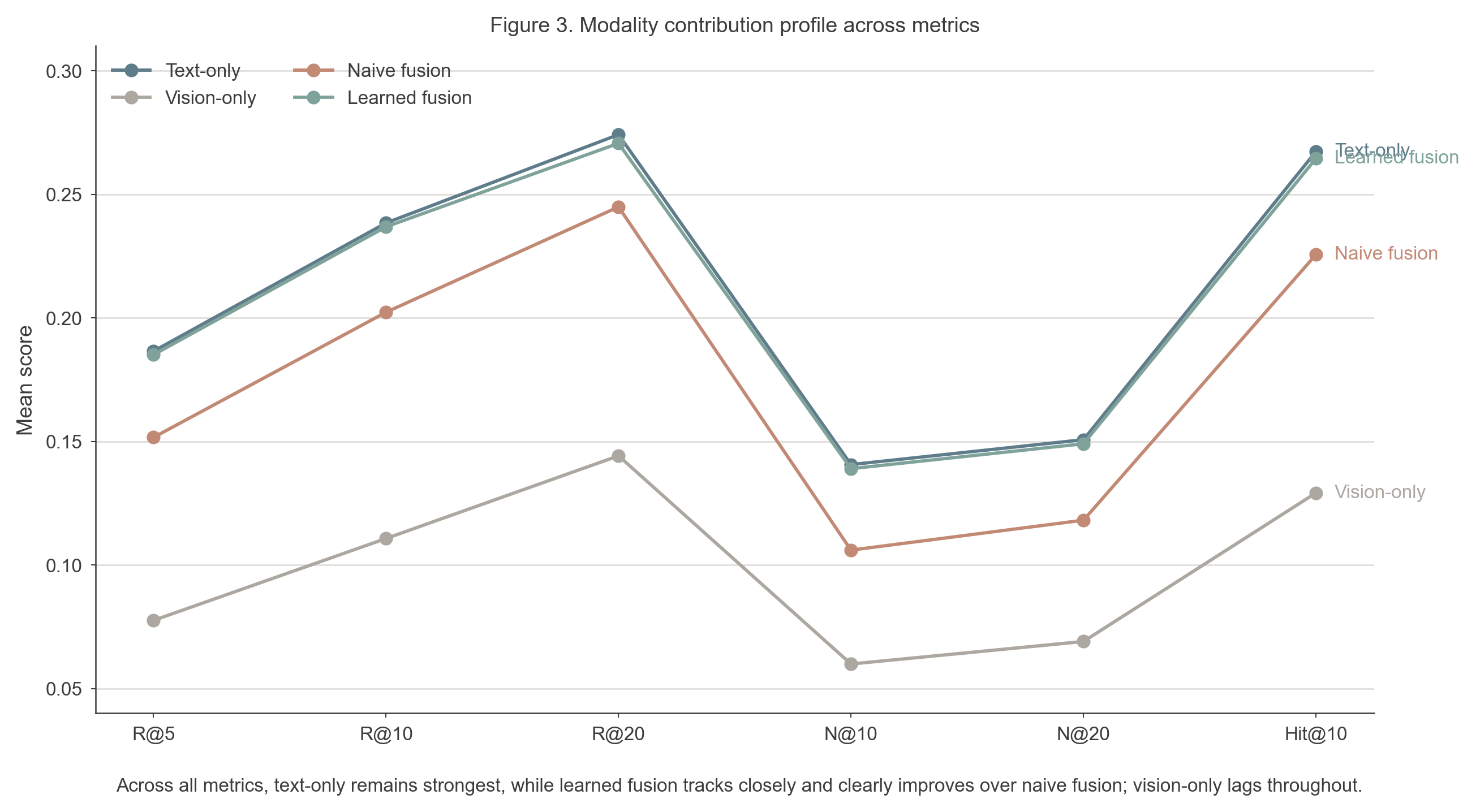}
  \caption{Modality contribution profile across metrics. Line plot showing performance trajectories for text-only, vision-only, naive fusion, and learned fusion strategies. Across all metrics, text-only remains strongest; learned fusion tracks closely and clearly improves over naive fusion, while vision-only lags throughout.}
  \label{fig:modality-profile}
\end{figure*}

A striking result in Table~\ref{tab:main-results} is that the strongest model is not a multimodal fusion architecture, but the text-only LLaVA representation itself.
On the LLaVA-covered subset, LLaVA text-only reaches 0.354 Recall@10, outperforming all fusion variants, including LLaVA + Attention (0.310), LLaVA + Concatenation (0.283), LLaVA + Naive Avg (0.281), and SMORE (LLaVA) (0.273).
The same ordering is visible in Figure~\ref{fig:modality-profile}, where the text-only curve remains consistently above all fusion strategies across the full set of metrics.
This result is important because it suggests that once visual content has been translated into semantically rich text, additional fusion with raw visual embeddings does not automatically provide complementary gains.

A plausible explanation is that LVLM-generated descriptions already compress much of the recommendation-relevant visual information into a highly usable semantic form.
Under this condition, adding CLIP-based visual features may introduce redundancy or noise rather than genuinely new signal, especially when the downstream retriever is intentionally simple.
The learned fusion methods still perform better than naive fusion, which shows that the fusion mechanism is not irrelevant.
However, even the best learned fusion variant cannot recover the gap to LLaVA text-only.
This indicates that, in our setting, the main bottleneck is not the lack of architectural sophistication, but the quality and usability of the underlying item representation.

\subsection{RQ3: How Does Representation Quality Affect Fusion Architectures?}
\label{sec:rq3}

To isolate the effect of representation quality from the effect of fusion architecture, we compare LLaVA-based and BERT-based variants under the same fusion mechanism.
The pattern is remarkably consistent.
For attention fusion, replacing BERT with LLaVA increases Recall@10 from 0.235 to 0.310.
For concatenation, the score rises from 0.189 to 0.283; for naive averaging, from 0.180 to 0.281; and for SMORE, from 0.188 to 0.273.
These gains are substantial across all four architectures, and Figure~\ref{fig:subset-heatmap} makes the pattern especially clear: the relative ordering is determined much more strongly by the representation source (LLaVA vs.\ BERT) than by the particular fusion operator applied on top of it.
This provides direct evidence that representation quality is the dominant factor in our setting.

The comparison also shows that stronger architectures cannot compensate for weak semantic inputs.
Among BERT-based methods, attention fusion yields only a small gain over BERT text-only, while concatenation, naive averaging, and SMORE all underperform the plain BERT baseline.
In contrast, once LVLM-generated descriptions are used, even simple fusion strategies become competitive, and all LLaVA-based variants consistently exceed their BERT-based counterparts.
Thus, the experimental evidence points to a clear conclusion: fusion architecture matters, but its benefit is secondary to the semantic richness of the item representation it operates on.
Improving representation quality produces large, architecture-agnostic gains, whereas changing the fusion mechanism alone yields comparatively limited returns.

\subsection{RQ4: Are the Improvements Consistent Across Metrics and Ranking Depths?}
\label{sec:rq4}

The advantage of LVLM-grounded representations is highly stable across both retrieval depths and evaluation metrics.
As shown in Table~\ref{tab:main-results}, LLaVA text-only improves over BERT text-only at every reported cutoff, achieving 0.289 vs.\ 0.198 at Recall@5, 0.354 vs.\ 0.228 at Recall@10, and 0.394 vs.\ 0.269 at Recall@20.
The same robustness holds for ranking-sensitive metrics: NDCG@10 improves from 0.184 to 0.274, and NDCG@20 from 0.198 to 0.285.
Figure~\ref{fig:subset-heatmap} further shows that this is not an isolated phenomenon tied to a single score or cutoff, but a broad and coherent pattern across the full evaluation suite.
Such consistency is particularly important because it suggests that the improvement is not limited to retrieving one or two easy hits near the top of the list, but reflects a more systematic enhancement of ranking quality.

Figure~\ref{fig:modality-profile} provides an additional robustness view by comparing the relative performance trajectories of different modality strategies across metrics.
The overall ordering remains stable: text-only is strongest, learned fusion follows, naive fusion trails behind, and vision-only consistently performs worst.
This stability indicates that the conclusions are not sensitive to the particular metric chosen.
Instead, the empirical picture is coherent across top-ranked retrieval, deeper candidate lists, and rank-aware evaluation.
From an experimental standpoint, this strengthens the claim that LVLM-grounded semantic representations provide a reliable improvement rather than a metric-specific artifact.

\subsection{RQ5: Why Do LVLM Descriptions Work So Well?}
\label{sec:rq5}

The quantitative results suggest that the advantage of LVLM-generated descriptions is systematic rather than incidental.
Compared with both short title embeddings and raw visual features, LVLM-grounded text provides a representation that is more directly usable for recommendation.
We attribute this advantage to three complementary properties.
First, LVLM descriptions encode \emph{compositional semantic richness}: a single description can jointly capture style, material, color, occasion, and functionality, whereas short titles are often too sparse and visual embeddings often remain entangled at the appearance level.
Second, they provide \emph{explicit contextual grounding}, making usage conditions such as formality, seasonality, or outfit compatibility directly available in the representation.
Third, they perform a form of \emph{implicit denoising} by focusing on product-relevant attributes and suppressing irrelevant visual artifacts such as background clutter, lighting, or packaging details.

These properties help explain the overall ranking pattern observed in Table~\ref{tab:main-results}.
If recommendation depends primarily on semantically meaningful item factors, then representations that explicitly encode such factors should be easier to match against user histories than either sparse titles or low-level visual embeddings.
This interpretation is consistent with the strong performance of LLaVA text-only, the weak performance of the vision-only baseline, and the limited gains from adding raw visual features on top of already strong LVLM-grounded semantics.
In this sense, the key benefit of LVLM descriptions is not simply that they are longer or more detailed, but that they transform visual evidence into a representation space that is substantially more recommendation-friendly.

\subsection{Qualitative Case Study}
\label{sec:qualitative}

To make the above mechanisms more concrete, we examine three representative cases in which LVLM-generated descriptions provide recommendation cues that are either missing from title-only representations or difficult to recover from raw visual features.

\paragraph{Case 1: Occasion-Aware Matching.}
A user who recently purchased formal accessories, including a pearl necklace, a black cocktail dress, and a silver bag, is recommended items for a subsequent formal-use scenario.
BERT text-only tends to recommend generic bracelets based largely on keyword overlap, without recognizing the formal context implied by the interaction history.
Vision-only retrieval instead overemphasizes superficial properties such as shininess or metallic appearance, which can lead to irrelevant suggestions.
In contrast, LLaVA-NeXT generates the description: ``Elegant silver-toned bracelet with crystal rhinestones, suitable for formal events and evening wear,'' which explicitly encodes occasion, style, and material.
This case illustrates how LVLM descriptions can recover recommendation-relevant event semantics that are largely absent from short titles and only weakly represented in visual embeddings.

\paragraph{Case 2: Style--Material Alignment.}
Consider a user whose recent history reflects a casual-wear preference, including a denim jacket, cotton T-shirt, and jeans.
BERT-based retrieval may return a mixture of footwear types, including formal shoes, because the title representation is too sparse to reliably encode the user's underlying style pattern.
Vision-only matching can also be unstable, often emphasizing local appearance similarity such as color or texture without capturing the broader semantic compatibility of the item.
LLaVA-NeXT instead produces the description: ``Casual canvas sneaker, durable canvas upper, relaxed fit, complements casual denim and cotton outfits,'' which explicitly links style and material cues.
This case shows that LVLM descriptions are able to bind material and stylistic information into a coherent semantic representation that better matches the user's preference profile.

\paragraph{Case 3: Seasonal Context.}
A third example involves a winter-oriented user whose recent purchases include a wool coat, knit beanie, and leather gloves.
BERT text-only tends to recommend generic scarves, including seasonally inappropriate variants, because the title signal is insufficient to capture seasonal usage.
Vision-only retrieval may focus on broad fabric-like appearance cues while failing to distinguish whether the item is appropriate for cold-weather wear.
LLaVA-NeXT generates the description: ``Cashmere scarf designed for cold weather protection, soft insulating material for winter wear, complements wool and knit accessories,'' which makes seasonal context and functionality explicit.
This example highlights that LVLM descriptions can surface pragmatic product semantics, such as weather suitability and layering compatibility, that are central to recommendation but difficult to infer reliably from native modality features alone.

\subsection{Limitations and Future Directions}
\label{sec:limitations}

While the results strongly support the effectiveness of \textbf{VLM4Rec}, several limitations should be acknowledged.
First, due to the computational cost of offline LVLM inference, we generated LLaVA descriptions for only 4,708 items, corresponding to 12.2\% of the full catalog.
As a result, the primary analysis for LVLM-grounded representations is conducted on the subset of 2,914 users whose training and evaluation interactions both fall within the covered item set.
Although this controlled design enables fair comparison across representation variants, it also means that the current study does not yet fully characterize the behavior of VLM4Rec under complete catalog coverage.
A natural next step is to scale semantic generation to the full item inventory and evaluate whether the observed gains remain stable in a fully deployed setting.

Second, our study focuses on a lightweight retrieval pipeline in order to isolate the effect of item representation quality.
This design is methodologically useful, but it leaves open the question of how VLM4Rec would interact with stronger downstream recommenders, such as sequential encoders, graph-enhanced user modeling, or learned ranking modules.
Future work could investigate whether LVLM-grounded multimodal semantic representations remain beneficial when integrated into more expressive recommendation architectures.

Third, the present framework uses LLaVA-NeXT 7B as the semantic grounding model primarily for efficiency reasons.
Larger or more specialized vision--language models may further improve semantic quality, but they would also introduce additional computational cost.
More broadly, our findings are currently grounded in a single domain and a single item coverage setting.
It therefore remains important to test VLM4Rec on broader datasets, domains, and model scales to better understand when LVLM-grounded semantic representations provide the largest advantage.

\section{Conclusion}
\label{sec:conclusion}

In this paper, we presented \textbf{VLM4Rec}, a lightweight framework that uses large vision--language models to construct multimodal semantic representations for recommendation.
Our results show that LVLM-grounded item representations consistently outperform traditional text features, raw visual embeddings, and several fusion-based alternatives in our setting.
Most notably, the best performance is achieved not by more complex fusion architectures, but by semantically rich text representations derived from item images.
These findings suggest that, in multimodal recommendation, improving representation quality can be more important than increasing fusion complexity.
We hope this work motivates further research on semantic representation as a first-class design principle for multimodal recommendation.

\balance
\bibliographystyle{ACM-Reference-Format}
\bibliography{references}

\end{document}